# Kelvin Probe Microscopy and Electronic Transport Measurements in Reduced Graphene Oxide Chemical Sensors

*Christopher E. Kehayias[1], Samuel MacNaughton[2], Sameer Sonkusale[2], and Cristian Staii[2,*]*

[1]Department of Physics and Astronomy and Center for Nanoscopic Physics, Tufts University, 4 Colby Street, Medford, MA 02155

[2]NanoLab, Department of Electrical and Computer Engineering, Tufts University, 200 Boston Avenue, Medford, MA, 02155

[*] CORRESPONDING AUTHOR: Cristian Staii, Department of Physics and Astronomy, Tufts University, 4 Colby Street, Medford, MA 02155, USA Ph.: (617) 627-5368.

E-mail: cristian.staii@tufts.edu

**Abstract**

Reduced Graphene Oxide (RGO) is an electronically hybrid material that displays remarkable chemical sensing properties. Here, we present a quantitative analysis of the chemical gating effects in RGO-based chemical sensors. The gas sensing devices are patterned in a field-effect transistor geometry, by dielectrophoretic assembly of RGO platelets between gold electrodes deposited on $SiO_2$/Si substrates. We show that these sensors display highly selective and reversible responses to the measured analytes, as well as fast response and recovery times (tens of seconds). We use combined electronic transport/Kelvin Probe Microscopy measurements to quantify the amount of charge transferred to RGO due to chemical doping when the device is exposed to electron-acceptor (acetone) and electron-donor (ammonia) analytes. We demonstrate that this method allows us to obtain high-resolution maps of the surface potential and local charge distribution both before and after chemical doping, to identify local gate-susceptible areas on the RGO surface, and to directly extract the contact resistance between the RGO and the metallic electrodes. The method presented is general, suggesting that these results have important implications for building graphene and other nanomaterial-based chemical sensors.

## 1. Introduction

Graphene and its chemical derivatives are proving to be very promising candidates for applications in a variety of fields, such as nanoscale electronics (1-4), mechanical engineering (5, 6), chemical sensing (7-10), and biosensing (11-14). Graphene has a characteristic two-dimensional honeycomb lattice structure (2), which makes the electronic transport of charge carriers in this material extremely sensitive to adsorption/desorption of gas molecules. Both bare and chemically functionalized graphene-based electronic devices are reported to be sensitive to various gases down to very low concentrations (6-9, 15-17), thus endorsing the use of graphene-related materials as nanoscale chemical vapor sensors.

Reduced Graphene Oxide (RGO) platelets constitute a very promising, cost-effective choice for building graphene-based electronic sensors (6, 9, 10, 15, 16, 18). RGO is an electronically hybrid material that can be controllably tuned from an insulator to a semiconductor material via reduction chemistry (6, 19, 20). Electrically conducting RGO platelets are typically obtained by exposing water-dispersed graphene oxide to different types of reducing agents, such as hydrazine (21), $NaBH_4$ (22), and ascorbic acid (16). The resulting RGO platelets are composed of carboxyl, alcohol, and dangling oxygen functional groups embedded within the familiar hexagonal lattice of carbon atoms (19). The availability of these functional groups allows RGO to interact with a wide range of chemical analytes, which act as either electron- donors or electron-acceptors on the sample surface, thus leading to a significant change in the resistance of the RGO-based device (15, 16, 19, 23). The RGO-based sensors are therefore amenable to chemical modifications that, in principle, permit to control their chemical sensitivity and selectivity to a very high degree. However, despite recent rapid progress there still are several important challenges that need to be dealt with, before RGO-based devices can be employed as versatile chemical and biological sensors. For example, detailed quantitative understanding of the analyte-RGO charge transfer, of the spatial distribution of the resulting charge carriers, and of the role played by the contact resistance between the RGO sample and the metal electrodes are essential for an optimal device design.

Here we combine electronic transport with Kelvin Probe Microscopy (KPM) experiments to probe variations in the surface potential and local charge distribution for RGO-based electronic sensors exposed to two types of chemical compounds: acetone ($(CH_3)_2CO$ which acts as an electron acceptor) and ammonia ($NH_3$, an electron donor). For these two analytes the variation of the device resistance and the induced charge are consistent with a chemical gating effect on the RGO conduction channel where hole conduction dominates. We quantify the amount of charge transferred to the sensor during chemical doping and spatially resolve the active sites of the sensor where the chemical gating process takes place. We also demonstrate that these experiments allow us to isolate the contributions from the contact resistance between the RGO and the metallic electrode, and show that the change in total resistance of our measured devices is mainly due to chemical doping of the RGO platelets.

## 2. Sample preparation and experimental setup

RGO platelets were prepared using a modified Hummers method as described previously (16, 23). Briefly, graphene oxide powder is obtained from graphite nanoplatelets via chemical treatment, and then further reduced using ascorbic acid (Vitamin C) treatment. The resulting RGO powder is suspended in dimethyl formamide, and then a fine nanoplatelet suspension is obtained via sonication. X-ray photoelectron spectra (XPS) and thermogravimetric analysis (TGA) performed on similar RGO platelets (16) demonstrate that the Vitamin C treatment is a mild and very effective alternative to hydrazine for reducing graphene oxide into RGO. To create assembly sites for the RGO platelets, Au electrodes were patterned on degenerately doped Si wafers containing a 200nm oxide layer using conventional sputtering techniques (23). RGO-based field effect transistors (RGO-FETs) (Figure 1a) were constructed by dielectrophoretic assembly of RGO platelets between the Au electrodes (23). The RGO network between the source and drain electrodes serves as conducting channels for the device. Atomic Force Microscopy measurements show that the RGO platelets have a typical linear size of 0.5-1μm and a typical thickness of 20-50nm. The gas sensing experiments for the RGO-FETs devices were performed at room temperature in a controlled environmental chamber (23). Initially dry, clean air is used as a reference gas and directed over the device at a rate of 0.5 lpm. Analyte gases (acetone or $NH_3$ at concentrations of 200 ppm (15, 16, 23)) are then sequentially substituted in the air flow (Figure 1) with the total flow rate held constant. Some of the samples were also exposed to $N_2$ flow at 0.5 lpm (instead of dry air). We find no significant difference between dry air and $N_2$ for the sensing response of the device (Supplementary Figure 1).

## 3. Results and discussion

### 3.1 Electronic transport and sensor response to chemical doping

Figure 1b shows the variation of the source-drain current $I_{SD}$ with the gate voltage $V_G$, for an RGO-FET device exposed to: dry air (red curve), acetone (blue curve) and $NH_3$ (green curve). The source-drain bias voltage was $V_B = 1V$ for all cases. The device shows p-dominated semiconducting behavior (i.e. hole conduction) over the whole range of applied gate voltages: $-40V \leq V_G \leq 40V$.

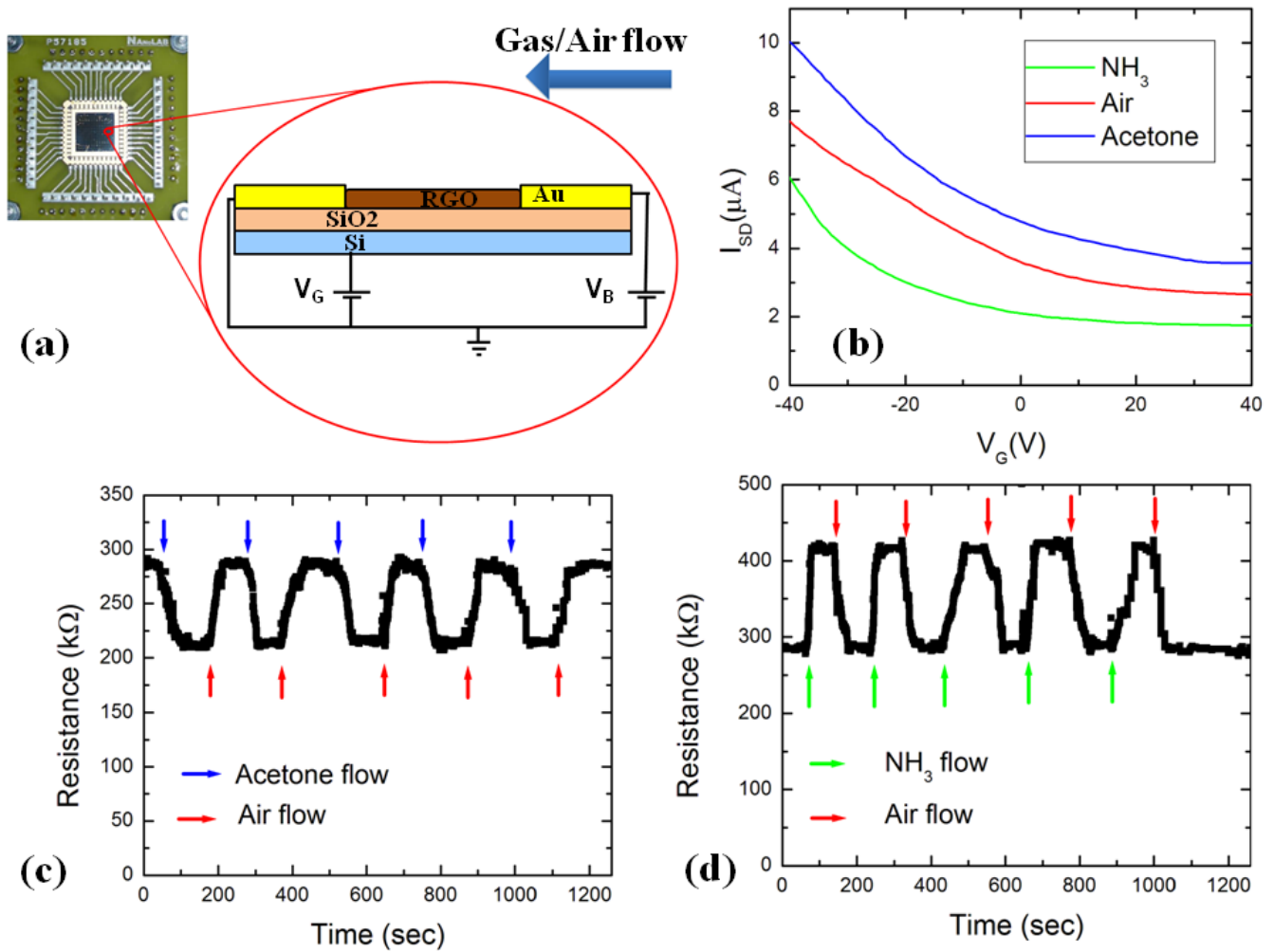

**Figure 1**. (a) Schematics of the electronic transport and gas sensing setup. The RGO-sensing devices are patterned on $SiO_2/Si$ wafer in a field-effect transistor configuration ($V_B$ and $V_G$ are the bias and gate voltages, respectively). RGO platelets are deposited between Au electrodes and aligned using dielectrophoresis. Each sensing device can be successively exposed to dry air and analytes (acetone and ammonia at a concentration of 200 ppm). (b) Source-drain current $I_{SD}$ versus gate voltage $V_G$ for a device while continuously exposed to: dry air (red curve), acetone (blue curve) and ammonia (green curve). The applied bias voltage was $V_B=1V$, and all gate sweeps were performed from -40V to +40V at a rate of 0.5V/sec. The device shows p-type response in air and the response to the analytes is characteristic for a chemical gating mechanism. (c) Dynamic response (device resistance versus time) for the device measured in (b), when exposed successively to dry air (red arrows) and acetone (blue arrows). 5 cycles are shown demonstrating reversibility and fast response and recovery times (tens of seconds). The decrease in device resistance upon acetone exposure is about 25%. (d) Dynamic response for the device measured in (b) and (c), when exposed successively to dry air (red arrows) and ammonia (green arrows). 5 cycles are shown demonstrating reversibility and fast response and recovery times (tens of seconds). The increase in device resistance upon exposure to ammonia is about 45%. The applied bias voltage was $V_B=1V$ and the gate voltage was $V_G = 0V$ for both (c) and (d).

The observed p-type behavior is consistent with previous results on RGO and graphene circuits obtained by chemical and thermal reduction of graphene oxide (15, 24, 25). The threshold voltage of the device (defined as the minimum gate voltage at which $I_{SD}$ is constant as a function of $V_G$) when exposed to air is $V_{th}^{air} = 23V$. When the device is exposed to acetone (an electron-withdrawing analyte (16, 23)) the source-drain current increases over the applied $V_G$ scan range and the threshold voltage shifts to a larger value: $V_{th}^{acetone} = 28V$ (blue curve in Figure 1b). Conversely, when the same device is exposed to $NH_3$ (an electron-donor analyte (15, 16)) we measure a decrease in $I_{SD}$ and a shift in the threshold voltage towards smaller values: $V_{th}^{NH3} = 16V$ (green curve in Figure 1b). These results demonstrate a charge transfer mechanism between the analyte (acetone or ammonia) and the RGO-FET sensor, where hole conduction dominates. $NH_3$ acts as an electron donor upon adsorbtion onto RGO (15), and it is expected to lower the concentration of charge carriers (holes), therefore decreasing the on-state current and shifting the threshold voltage towards the negative regime. Acetone serves as electron acceptor (16, 23), and it is therefore expected to increase the charge carrier concentration, which results in increased on-state currents and threshold voltages. Similar chemical doping mechanisms has previously been proposed for graphene and RGO based sensors (7, 8, 15, 17).

The shifts of the $I_{SD}$-$V_G$ transport curves in Figure 1 can be used to calculate the relative change in the density of charge carriers for acetone ($\Delta n_a$) and $NH_3$ ($|\Delta n_{NH3}|$) doping. Using $e \cdot \Delta n = C \cdot \Delta V_{th}$, where $e$ is the electron charge, $C$ is the $SiO_2$ gate capacitance and $\Delta V_{th}$ is the shift of the threshold voltage when the device is exposed to the analyte, measured relative to the threshold voltage when the device is exposed to air (15, 26), we obtain:

$$\frac{\Delta n_a}{|\Delta n_{NH3}|} = \frac{V_{th}^{aacetone} - V_{th}^{air}}{V_{th}^{air} - V_{th}^{NH3}} \approx 0.7 \qquad (1)$$

In Figure 1c, d we show the dynamic response (device resistance vs. time) for the RGO-FET sensor when exposed to analytes (acetone in Figure 1c, and $NH_3$ in Figure 1d, respectively). The resistance of the sensor was measured for $V_B= 1V$ and $V_G= 0V$ first in air, then when the device was periodically exposed to analyte, and then air. The duration of each air (or analyte) exposure was about 100 sec. Five complete exposure cycles are shown in Figure 1c (acetone) and Figure 1d ($NH_3$) to demonstrate the reproducibility and the stability of the sensing response. Upon exposure to acetone the sensor resistance *decreased* by about 25% (Figure 1c), while when the device was exposed to $NH_3$ its resistance *increased* by about 45% (Figure 1d). The sign of the dynamic response (decrease in device resistance for acetone, respectively increase in resistance for $NH_3$) are again consistent with the expected chemical gating response for the electron-withdrawing acetone and the electron-donor $NH_3$ vapors, which lead to an increase (decrease) in the concentration of charge carriers in the p-type RGO device. Figure 1c and d also show that the sensor response to the analytes and its recovery upon air exposure are rapid (tens of seconds) and reversible for the relatively low analyte vapor concentrations (200 ppm) studied here. We note that the overall sensing performance (response and recovery times, sensitivity and stability of the sensor) for these devices is at the level of the state of the art performance for graphene-based sensors reported in literature (10, 15, 18). This demonstrates that both the reduction method (based on Vitamin C) and the dielectrophoretic assembly of RGO platelets are viable, high-throughput techniques for fabricating low-cost, high-sensitivity chemical sensors.

*3.2 Kelvin Probe Microscopy of RGO devices*

To gain a deeper quantitative understanding of the sensing mechanism we perform Kelvin Probe Microscopy (KPM) measurements of RGO devices both in air and upon exposure to acetone and $NH_3$. KPM is a dual-pass AFM technique (27-29). In the first line scan the AFM-tip acquires the topography profile of the sample in tapping mode (Figure 2a). In the second line scan, the tip travels at a preset height above the sample surface (h=30 nm for the experiments presented here), as shown in Figure 2b. During the second pass the AFM cantilever is excited electrically by applying both a *dc* voltage $V_{dc}$, and an *ac* component (with amplitude $V_{ac}$= 1V and frequency equal to the resonant the frequency of the cantilever oscillation ($\omega_0 \approx 75$KHz, in our experiments). The *dc* component is controlled by a feedback loop such that the amplitude of the cantilever at $\omega_0$ equals zero. Under these conditions the dc tip voltage $V_{dc}$ matches the surface potential $\Phi(x,y)$ beneath the tip (27-29). Thus by mapping $V_{dc}$ versus the tip position, the KPM image records the distribution of the surface potential $\Phi(x,y)$ along the sample surface with very high spatial resolution (Figures 3, 4). Assuming that the work function of the tip and that of the charge-neutral RGO are constant, we have that $V_{dc}$ directly tracks changes in the surface potential resulting from variations in the local charge carrier concentration. We note that a similar approach was used in reference (28) for mapping variations in the Fermi energy of charge neutral single-layer graphene on SiC (0001) surfaces. For the measurements reported here KPM images were taken on an Asylum Research MFP3D AFM using platinum – coated tips with curvature radius R=20-30nm, quality factor Q=150 and spring constant k=0.65-1 N/m.

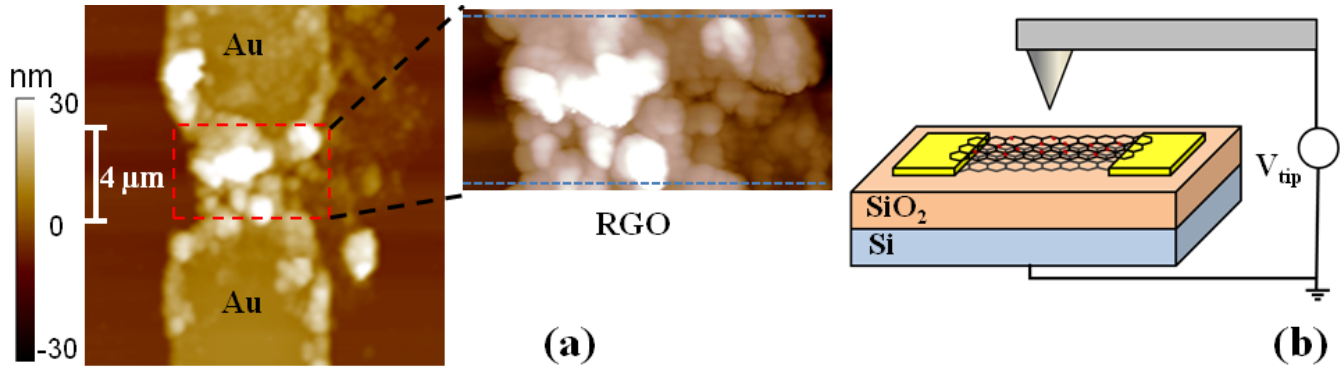

**Figure 2.** (a) AFM topographic image of an RGO-sensing device. The two Au electrodes (source and drain) and the RGO platelets are shown in the image. The RGO platelets are assembled in the central region, between the two electrodes. The RGO platelets have a typical linear size of 0.5-1μm and a typical thickness of 20-50nm, as characterized by AFM. The red dotted square marks the region between the electrodes used for the KPM measurements shown in Figures 3 and 4. This region is shown at higher resolution in the middle inset. The blue dotted lines in the inset mark the position of the Au electrodes. (b) Schematics of the Kelvin Probe Microscopy measurement, with overlapping RGO platelets assembled between the Au electrodes (see main text).

Figure 2 shows a typical AFM-topography image of a sample, including the two electrodes (Figure 2a), and the schematics of the KPM setup (Figure 2b). We have performed KPM measurements over the RGO regions between the two electrodes (indicated by the dotted red square, and shown as an inset in Figure 2a), with the sample exposed to air (Figure 3a), acetone (Figure 3b) and then again to air (Figure 3c). This sequence corresponds to a complete exposure cycle shown in transport data (Figure 1c). The corresponding distributions in the surface potential are shown respectively in Figure 3d, e and f. The KPM data shows a clear shift in the average distribution of the surface potential upon exposure to acetone from $V_{air} = (172.7 \pm 33) mV$ (Figure 3d) to $V_{acetone} = (242 \pm 38) mV$ (Figure 3e). After exposure to acetone the sample is exposed again to air and the average of the surface potential distribution $V_{air}^{after} = (176.2 \pm 25) mV$ approaches the initial average of the distribution.

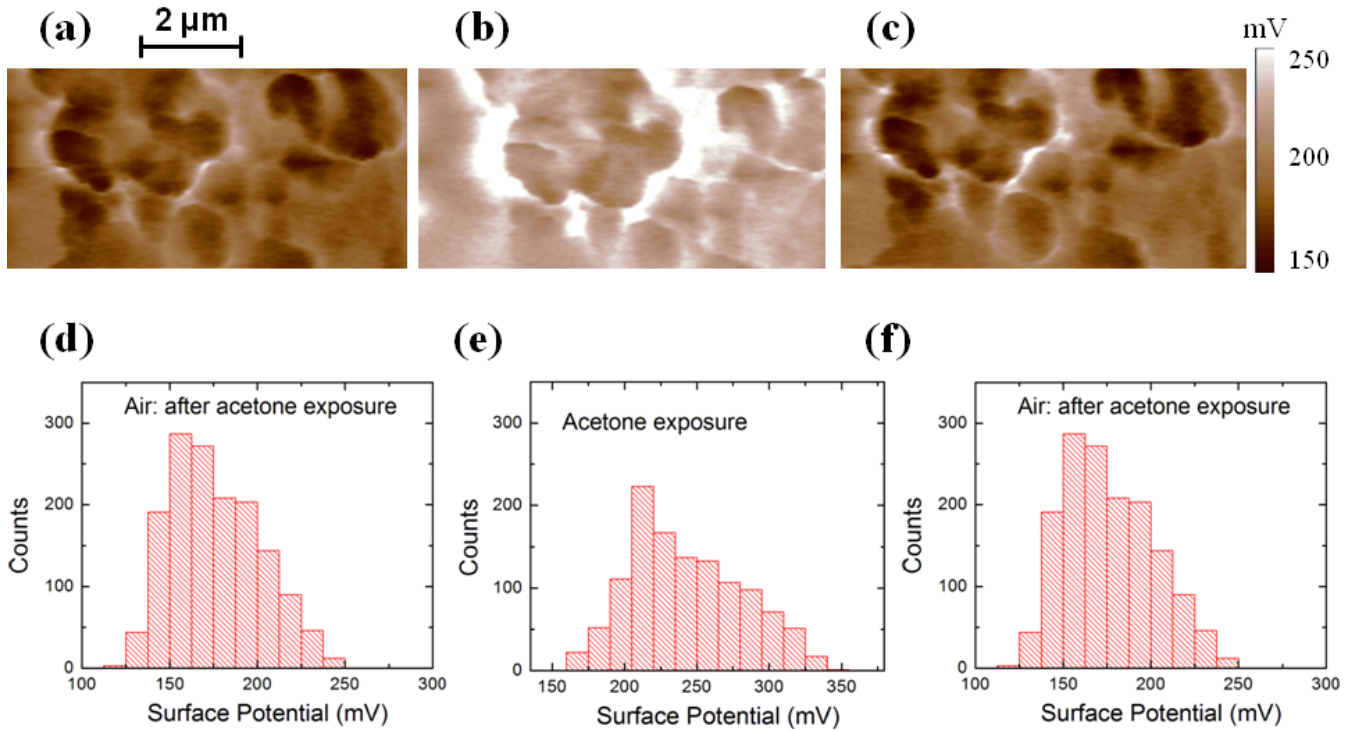

**Figure 3.** (a)-(c) KPM potential maps of an RGO sensor when exposed to: dry air (a), acetone (b) and dry air (c). Figure (b) is taken while the acetone is applied (the sample was exposed to acetone for 15 min prior to the measurements, and then continuously exposed to acetone during the KPM experiments). Figure (c) is taken in dry air, and after the sample was continuously exposed to dry air for 15 min following the acetone exposure. The images show the variation in the surface potential due to gas exposure, and the active binding regions where chemical doping takes place. (d)-(f) Histograms of surface potential for the images shown in (a)-(c), respectively. The average values/standard deviations for these distributions are: $V_{air} = (172.7 \pm 33) mV$, from (d), $V_{acetone} = (242 \pm 38) mV$, from (e), and $V_{air}^{after} = (176.2 \pm 25) mV$, from (f) (the quoted experimental uncertainties are the standard deviations of the distributions). The distribution for the surface potential obtained during exposure to acetone are significantly different from the distributions obtained in air ($p \leq 0.05$, one way ANOVA). These distributions are used to quantify the amount of charge transferred during chemical doping.

The chemical doping model implies that the observed shifts in the surface potential are due to charge transfer between the analyte (acetone) and the sample: vapor molecules of acetone adsorb near the sample surface, thus changing the electrostatic environment and increasing the concentration of hole charge carriers. This leads to an increase surface potential as measured by KPM. We note that the KPM images have very high spatial resolution and thus permit to identify the regions of the sample where the chemical doping effect takes place, as well as the local changes in the surface potential due to charge transfer between the analyte and the RGO sample. For example, the comparison between Figure 3a and 3b shows that the chemical doping is not uniform along the RGO surface, and also permits to identify the gate-susceptible regions on the sensor surface (i.e. regions where the change in surface potential is maximum upon doping). To our knowledge these represent the first high-resolution measurements of charge transfer in RGO gas sensors.

*3.3 Analysis of the combined Kelvin Probe Microscopy/electronic transport measurements*

We combine the KPM and transport data to quantify the amount of charge transferred due to chemical doping, as well as the contact resistance between the RGO and the Au electrodes. Assuming negligible change in the total sample capacitance $C$ during acetone exposure we have that: $C = \frac{Q_{air}}{V_{air}} = \frac{Q_{air} + \Delta q_a}{V_{acetone}}$, where $Q_{air} = e \cdot A \cdot n_{air}$ is the total electric charge when the device is exposed to air, and $\Delta q_a = e \cdot A \cdot \Delta n_a$ is the increase in the electric charge due to acetone exposure (here $A$ is the sample surface area, while $n_{air}$ and $\Delta n_a$ represent the density of charge carriers in air, and the increase in the density of charge carriers, respectively). From this expression it follows that: $\Delta q_a = Q_{air} \cdot \frac{V_{acetone} - V_{air}}{V_{air}}$, or:

$$\frac{\Delta n_a}{n_{air}} = \frac{V_{acetone} - V_{air}}{V_{air}} \quad (2)$$

Equation 2 shows that the percentage change in the charge carrier density due to doping is equal to the percentage change in the surface potential measured by KPM. Using the data shown in Figure 3 d, and c we obtain: $\frac{\Delta n_a}{n_{air}} = 0.4 \pm 0.2$, that is an increase in the surface charge density by about 40% (from the average of the two surface potential distributions shown in Figure 3). The absolute values for $n_{air}$ and $\Delta n_a$ can be estimated from the above result and the $I_{SD}$-$V_G$ transport curves (Figure 1b). Assuming a parallel plate model for the total capacitance $C$ we have (15, 24): $\Delta n_a = \left(\frac{\varepsilon_0 \cdot \varepsilon}{t \cdot e}\right) \cdot \Delta V_{th}$, where $\varepsilon_0$ and $\varepsilon$ are the permittivity of the free space and of the SiO$_2$, and $t$=200 nm is the thickness of the SiO$_2$ layer. With $\Delta V_{th} = 5V$ (Figure 1b) and $\Delta n_a / n_{air} \approx 0.4$ (see above) we obtain: $\Delta n_a \approx 5 \cdot 10^{11} cm^{-2}$ and $n_{air} \approx 1.2 \cdot 10^{12} cm^{-2}$, values that are similar to typical charge carrier densities found for RGO and graphene- based sensors (3, 6, 17, 19).

Next, we combine the KPM results with the results obtained from the dynamic response (Figure 1c) to quantify the contact resistance between the RGO and the electrode. The chemical doping model assumes that the surface coverage $\theta_{RGO}$ of analyte molecules on RGO (which is proportional to the available binding sites on RGO surface) determines the change in sample conductance and it is given by (15, 30):

$$\delta \cdot \theta_{RGO} = \frac{\Delta n_a}{n_{air}} \quad (3)$$

where $\delta$ is a proportionality constant, for a given analyte concentration and exposure time (15). As in reference (15) we find that under these general assumptions, the change in the absolute value of the total resistance of the RGO platelets, $|\Delta R_{RGO}|$ due to gas exposure is:

$$\frac{|\Delta R_{RGO}|}{R_{RGO}} = \frac{|\Delta R - \Delta R_C|}{R - R_C} = \frac{\delta \cdot \theta_{RGO}}{1 + \delta \cdot \theta_{RGO}} = B \quad (4)$$

where $R_{RGO}$ is the resistance of the RGO platelets in air, $R$ is the total resistance of the RGO device in air, $R_C$ is the sample contact resistance in air, and $\Delta R$ and $\Delta R_C$ are respectively, the changes in the total sample resistance and the contact resistance due to analyte exposure. Equation 4 shows that the relative change in the resistance of the RGO platelets is a constant ($B$) for given analyte concentration and exposure time. By combining equations (3) and (4) above and using our result from KPM measurements: $\Delta n_a / n_{air} \approx 0.4$ we obtain:

$$B = \frac{\left(\dfrac{\Delta n_a}{n_{air}}\right)}{1 + \left(\dfrac{\Delta n_a}{n_{air}}\right)} \approx 0.29 \qquad (5)$$

In addition, from the dynamic response data (Figure 1c) we have that:

$$\frac{|\Delta R|}{R} \approx 0.25 \qquad (6)$$

Furthermore, our KPM data (Figure 3a-c) also show no significant change in the surface potential near the electrodes during acetone exposure, implying a negligible change in contact resistance upon doping: $\Delta R_C / R \approx 0$. This result is usually assumed in literature (see for example ref. (15)), however to our knowledge, this is the first direct measurement which shows that the change in the sensor contact resistance is very small, during chemical doping. This is a non-trivial result demonstrating that that the sensing response is mainly due to charge transfer between RGO and analytes, and is not due to Schottky contacts or functional groups at the interface between the RGO and the electrodes. Combining this result with the results given by equations (4), (5) and (6) we conclude that the contact resistance between the RGO platelets and the sample electrode is negligible compared to the total sample resistance, that is:

$$\frac{R_C}{R} \approx 1 - \frac{1}{B} \cdot \left(\frac{|\Delta R|}{R}\right) \approx 0.14 \qquad (7)$$

Since: $R = 287 K\Omega$ and $|\Delta R|/R = 0.25$ (Figure 1c), equation (7) gives that: $R_C \approx 40 K\Omega$. We emphasize that this result is obtained directly from measurements, by combining KPM and transport data. Thus the combination of KPM and transport/dynamic resistance data provides a general method for measuring the contribution of the contact resistance between the RGO sample and the metallic electrodes. This has important practical implications for building chemical sensors as the contact resistance could distort the sensitivity of the sensing device (15).

Finally, we note that similar results are obtained from the data on $NH_3$ sensing, thus demonstrating the consistency of the described combined KPM-transport method. Figure 4 shows the KPM image (Figure 4a) and surface potential distribution (Figure 4b) for the device exposed to $NH_3$. From the distribution of the surface potential we get the average value $V_{NH3} = (60 \pm 40) mV$, that is a negative shift with respect to $V_{air} = (178 \pm 31) mV$, consistent with a decrease in surface charge density. Combining this data with the dynamic response measurements (Figure 1d), and following the same analysis as for acetone we obtain: $\dfrac{|\Delta n_{NH3}|}{n_{air}} \approx 0.6$, and negligible contact resistance, $R_C < 10 K\Omega$. This value is consistent with the corresponding $R_C$ found from the analysis of the acetone data (see above), due to the relatively large standard deviation obtained for the $NH_3$ surface potential distribution (Figure 4b). The KPM data allows us to estimate the ratio between the changes in carrier concentrations due to acetone vs. $NH_3$ doping: $\dfrac{\Delta n_a}{|\Delta n_{NH3}|} \approx \dfrac{0.4}{0.6} = 0.66$. This result is in excellent agreement with the result given by equation 1, which was obtained independently from the $I_{SD}$-$V_G$ transport curves. Finally, by combining the KPM and transport data we obtain a level of doping (decrease in hole concentration) due to $NH_3$ exposure of: $|\Delta n_{NH3}| \approx 7 \cdot 10^{11} cm^{-2}$.

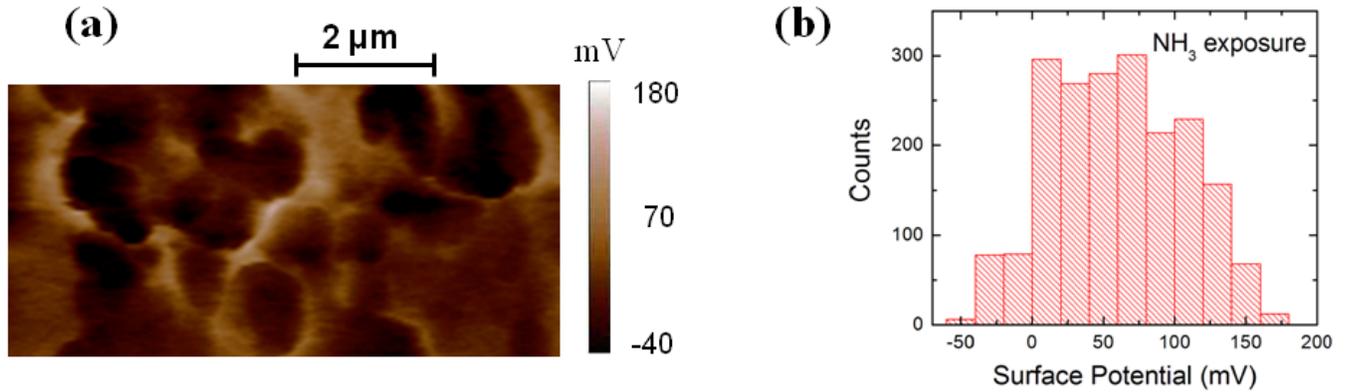

**Figure 4.** (a) KPM potential maps of an RGO sensor when exposed to ammonia. The image shows the variation in the surface potential due to gas exposure, and the active binding regions where chemical doping takes place. (b) Histogram of the surface potential for the image shown in (a). The average value/standard deviation for this distribution is $V_{NH3} = (60 \pm 40) mV$, consistent with a decrease in the carrier charge density due to adsorption of the electron-donor ammonia vapors.

## 4. Conclusions

We have used a combined electronic transport/Kelvin Probe Microscopy technique to report the first high-resolution measurements of chemical doping effects in reduced graphene oxide based gas sensors. The sensing devices are obtained by dielectrophoretic assembly of RGO platelets between Au electrodes, and show high specificity and reversibility when exposed to acetone and ammonia. KPM allows us to visualize the active binding regions where chemical doping takes place, and the electronic transport measurements (dynamic response, and current versus gate voltage transport curves) provide the global sensor response to gaseous analytes. Together, these two types of measurements show that the doping occurs mainly within the RGO platelets with negligible contribution from contact resistance. We also quantify the amount of charge transferred to RGO due to the chemical doping when the device is exposed to electron acceptor (acetone) and electron donor (ammonia) analytes, and measure directly the contact resistance between the RGO and the metallic electrodes. Here we have analyzed the global sensing effect of the entire RGO sample. However the analysis method could be extended to study local chemical gating effects within smaller gate-susceptible regions of the RGO sample, and to correlate the local surface charge with surface functional groups on RGO. These measurements could also be extended to measure changes in contact resistance and Schottky effects due to chemical doping in graphene-based sensors. The method presented is general and can be applied to any 2-dimesnional chemical sensor, where the sensing mechanism is based on the charge transfer between the analyte and the sensor.

**Acknowledgments**

We thank Prof. Sanjeev K. Manohar (U-Mass Lowell) for providing the RGO platelets used for patterning the chemical sensors. This work was supported by the Tufts Faculty Research Award (FRAC) and the Tufts Summer Scholar Program (CK).

Supplementary Figures

**Kelvin Probe Microscopy and Electronic Transport Measurements in Reduced Graphene Oxide Chemical Sensors**

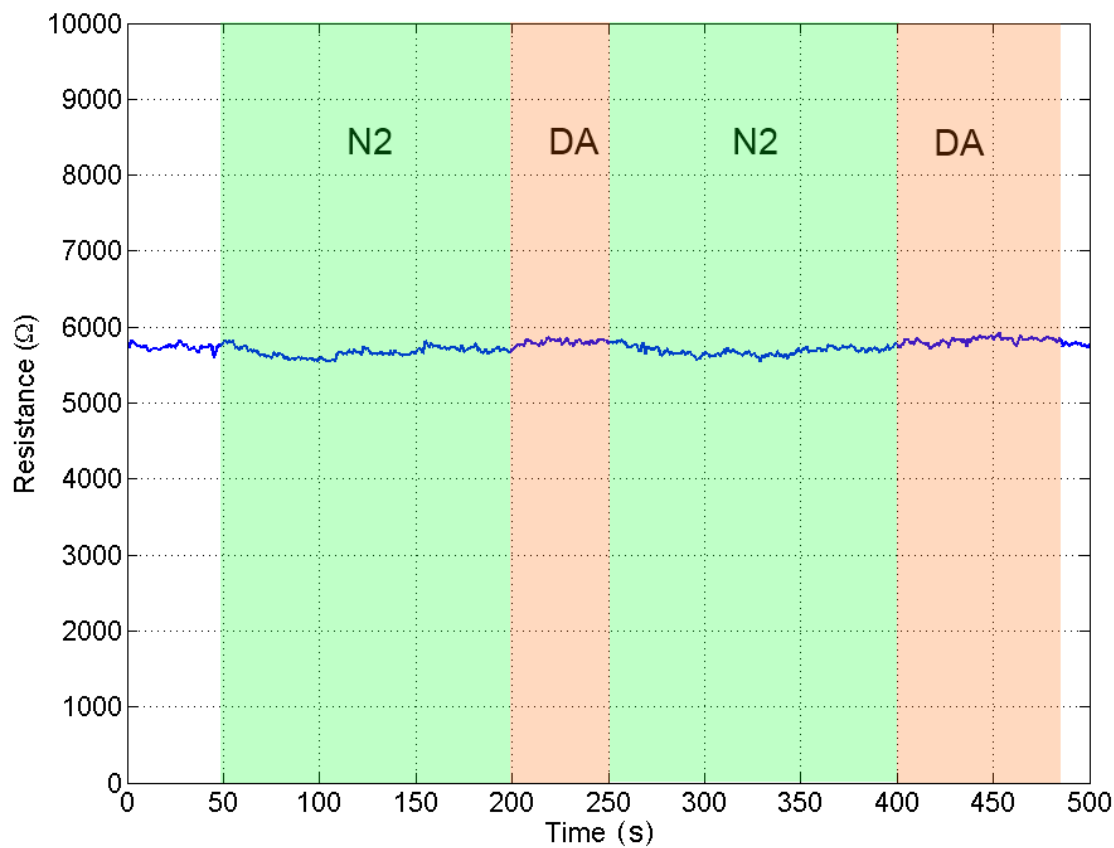

**Figure S1.** Device resistance versus time for an RGO device, when exposed successively to dry air (DA) and nitrogen gas (N2). There is no measurable change in resistance when the device is exposed to nitrogen compared to dry air.